\documentclass[a4paper,12pt]{article}

\pdfoutput=1

\usepackage{amsmath}
\usepackage{amssymb}
\usepackage{mathrsfs}
\usepackage{bbm}
\usepackage{graphicx,subfigure}
\usepackage[numbers,sort&compress]{natbib}

\usepackage{color}
\usepackage{ulem}

\numberwithin{equation}{section}

\newlength{\dinwidth}
\newlength{\dinmargin}
\newcommand{\abs}[1]{\left\lvert #1 \right\rvert}

\begin{document}

\title{\bf \Large
Phenomenological discriminations of the Yukawa interactions in two-Higgs doublet models with $Z_2$  symmetry }
\author{Xiao-Dong Cheng\footnote{chengxd@iopp.ccnu.edu.cn}, Ya-Dong Yang\footnote{yangyd@mail.ccnu.edu.cn} and Xing-Bo Yuan\footnote{xbyuan@mails.ccnu.edu.cn}\\
{\small Institute of Particle Physics, Central China Normal University, Wuhan, Hubei 430079, P.~R.~China}\\
{\small Key Laboratory of Quark \& Lepton Physics, Ministry of Education, P.~R.~China}}

\bigskip
\date{}
\maketitle

\begin{abstract}
{\noindent}There are four types of  two-Higgs doublet models under a
discrete $Z_2$ symmetry imposed to avoid tree-level flavour-changing
neutral current, i.e. type-I, type-II, type-X and type-Y models. We
investigate the possibility to discriminate the four models in the
light of the flavour physics data, including $B_s-\bar B_s$ mixing,
$B_{s,d} \to \mu^+ \mu^-$, $B\to \tau\nu$ and $\bar B \to X_s \gamma$
decays, the recent LHC Higgs data, the direct search for charged Higgs
at LEP, and the constraints from perturbative unitarity and vacuum
stability. After deriving the combined constraints on the Yukawa
interaction parameters, we have shown that the correlation between the
mass eigenstate rate asymmetry $A_{\Delta\Gamma}$ of  $B_{s} \to \mu^+
\mu^-$ and the ratio $R={\cal B}(B_{s} \to \mu^+ \mu^-)_{\rm exp}/
{\cal B}(B_{s} \to \mu^+ \mu^-)_{\rm SM}$ could be a sensitive probe to  discriminate the four models
with future precise measurements of the observables in the $B_{s} \to \mu^+ \mu^-$ decay at LHCb.

\end{abstract}
\newpage
\section{Introduction}
Although the Standard Model (SM) for particle physics has been successful for over three decades, it still shows some problems which solutions could imply physics beyond its scope~\cite{Ref:weinberg:1,Ref:weinberg:2,Ref:weinberg:3,Ref:weinberg:4,Ref:weinberg:5,Ref:neutrio,Ref:baryonasymmetry:1,Ref:baryonasymmetry:2,Ref:baryonasymmetry:3,Ref:baryonasymmetry:4}. Recently, the ATLAS~\cite{ATLAS:1,ATLAS:2} and CMS~\cite{CMS:1,CMS:2} experiments at LHC have discovered a new neutral boson with properties consistent with those of the SM Higgs boson~\cite{Higgs:64:1,Higgs:64:2,Higgs:66,Englert:1964et,Guralnik:1964eu,Kibble:1967sv}. With the experimental progress at LHC, it is of great interest to confirm whether this boson is the only one fundamental scalar just as the SM,   or belongs to an extended scalar sector responsible to the electroweak symmetry breaking (EWSB). The simplest scenario entertaining the latter possibility is provided by the two-Higgs doublet models (2HDM).

Besides the SM Higgs sector, an additional Higgs doublet is introduced in the 2HDMs. This class of  models can provide new source of CP violation beyond the SM~\cite{Ref:2hdmcpv}, which are needed to explain the observed cosmic matter-antimatter asymmetry. The 2HDMs could also be understood as an effective theory for many natural EWSB scenarios, such as the Minimal Supersymmetric Standard Model (MSSM) \cite{Ref:THDMreport}.

However, unlike the SM, the tree-level flavour-changing neutral
current (FCNC) transition in the 2HDM is not forbidden by the
Glashow-Iliopoulos-Maiani (GIM) mechanism. These FCNCs can cause
severe phenomenological
difficulties~\cite{McWilliams:1980kj,Shanker:1981mj,Hou:1991un}. Besides
some other solutions \cite{MFV:1,MFV:2,MFV:3,MFV:2HDM:1,MFV:2HDM:2},
this problem can be addressed by imposing a discrete $Z_{2}$
symmetry~\cite{Ref:GW}. According to different $Z_2$ charge
assignments, there are four types of 2HDMs, referred to, respectively, as the type-I, type-II, type-X and type-Y 2HDMs~\cite{Ref:mayumiaoki}. Therefore, phenomenologically distinguishing between these 2HDMs is an important issue and worthy of detailed investigation~\cite{Rindani:2013mqa}.

The 2HDMs present very interesting phenomena in both low-energy flavour transitions such as $B\to X_s \gamma$ decay and $B_s-\bar B_s$
mixing, and high-energy collider processes such as various Higgs decay channels. At present, many analyses have been performed \cite{2HDM:ph:1,2HDM:ph:2,2HDM:ph:3,2HDM:ph:4,2HDM:ph:5,2HDM:ph:6,2HDM:ph:7,2HDM:ph:8,2HDM:ph:9,2HDM:ph:10,2HDM:ph:11,2HDM:ph:12,2HDM:ph:13,2HDM:ph:14,2HDM:ph:15,2HDM:ph:16,2HDM:ph:17}, however, most of them concentrate on the type-II 2HDM. In this work, we shall extend the previous analyses and study the possibility to discriminate   the four different types of 2HDM in favor of experimental measurement. To constrain the model parameters, we shall consider the following constraints:
\begin{itemize}
\item flavour processes: $B_s-\bar B_s$ mixing, $\bar B \to X_s\gamma$, $B\to \tau\nu$ and $B_{s,d}\to \mu^+\mu^-$ decays,
\item direct search for Higgs bosons at LEP, Tevatron, and LHC,
\item perturbative unitarity and vacuum stability.
\end{itemize}
For the $B_s \to \mu^+ \mu^-$ decay, there are several interesting observables very sensitive to new physics effects as suggested recently by \textit{De Bruyn et al}.~\cite{DeBruyn:2012wk}. In this paper, we use these observables to probe the 2HDMs and find the correlation between  the mass eigenstate rate asymmetry $A_{\Delta\Gamma}$  and the ratio $R={\cal B}(B_{s} \to \mu^+ \mu^-)_{\rm exp}/ {\cal B}(B_{s} \to \mu^+ \mu^-)_{\rm SM}$, which could be used to  discriminate the four models with future precise measurements of the observables in the $B_{s} \to \mu^+ \mu^-$ decay at LHCb.

Our paper is organized as follows: In the next section, we give a
brief review on the 2HDM with the $Z_2$ symmetry. In
section~\ref{sec:flavourobservable}, the theoretical formalism for the
flavour observables are presented. In
section~\ref{sec:numericalanalysis}, we give our detailed numerical
results and discuss the possibility of discriminating the four types
of 2HDM. Our conclusions are given in section~\ref{sec:conclusion}. The relevant Wilson coefficients due to the contributions of 2HDMs are presented in the appendices~\ref{appendix:1} and~\ref{appendix:2}.

\section{2HDM under the $Z_2$ symmetry}
\label{sec:formalism}
In the 2HDM,  the two Higgs doublets $\Phi_{1}$ and $\Phi_{2}$ can be generally parameterized as
\begin{align}
\Phi_i=
\begin{pmatrix}
\omega_i^+\\
\frac1{\sqrt2}(v_i+h_i-iz_i)
\end{pmatrix}.
\end{align}
For a CP-conserving Higgs potential, the two vacuum expectation values (vevs) $v_1$ and $v_2$ are real and positive ~\cite{Ref:THDMreport}. They satisfy the relations $v_1=v\cos\beta $ and $v_2=v\sin\beta$ with $v = 246\,\rm GeV$. The physical scalars can be obtained by the rotations
\begin{align}
\begin{pmatrix}h_1\\h_2\end{pmatrix}=R(\alpha)
\begin{pmatrix}H\\h\end{pmatrix},\quad
\begin{pmatrix}z_1\\z_2\end{pmatrix}=R(\beta)
\begin{pmatrix}z\\A\end{pmatrix},\quad
\begin{pmatrix}\omega_1^+\\\omega_2^+\end{pmatrix}=R(\beta)
\begin{pmatrix}\omega^+\\H^+\end{pmatrix},
\end{align}
where the rotation matrix is given by
\begin{align}
R(\theta)=\begin{pmatrix}\cos\theta&-\sin\theta\\
\sin\theta&\cos\theta\end{pmatrix}.
\end{align}
The mixing angles $\alpha$ and $\beta$ are determined by the parameters of the Higgs potential. The physical Higgs spectrum consists of five degrees of freedom: two charged scalars $H^\pm$, two CP-even neutral scalars $h$ and $H$, and one CP-odd neutral scalar $A$.

In the interaction basis, the Yukawa interactions of these Higgs bosons can be written as
\begin{align}
- \mathcal L_Y=
\bar Q_L( Y_1^d\Phi_1+Y_2^d\Phi_2) d_R
+\bar Q_L(Y_1^u \tilde\Phi_1+Y_2^u \tilde\Phi_2)u_R
+\bar L_L(Y_1^\ell\Phi_1+Y_2^\ell\Phi_2) e_R
+\text{H.c.},
\end{align}
where $\tilde\Phi_i=i\sigma_2\Phi_i^*$ with $\sigma_2$ the Pauli
matrix, $Q_L$ and $L_L$ denote the left-handed quark and lepton
doublets, and $u_R$, $d_R$ and $e_R$ are the right-handed up-type quark, down-type quark and lepton singlet, respectively. The Yukawa coupling matrices $Y_i^f$ ($f=u,d,\ell$) are $3\times3$ complex matrices in flavour space.
\begin{table}[t]
\centering
\begin{tabular}{|l||c|c|c|c|c|c|}
\hline & $\Phi_1$ & $\Phi_2$ & $u_R$ & $d_R$ & $\ell_R$ & $Q_L$, $L_L$ \\  \hline
Type-I  & $+$ & $-$ & $-$ & $-$ & $-$ & $+$ \\
Type-II & $+$ & $-$ & $-$ & $+$ & $+$ & $+$ \\
Type-X  & $+$ & $-$ & $-$ & $-$ & $+$ & $+$ \\
Type-Y  & $+$ & $-$ & $-$ & $+$ & $-$ & $+$ \\
\hline
\end{tabular}
\caption{\small Charge assignments of the $Z_2$ symmetry in the four types of 2HDM.} \label{Tab:type}
\end{table}

\begin{table}[t]
\centering
\begin{tabular}{|l||c|c|c|c|c|c|c|c|c|}
\hline
& $\xi_h^u$ & $\xi_h^d$ & $\xi_h^\ell$
& $\xi_H^u$ & $\xi_H^d$ & $\xi_H^\ell$
& $\xi_A^u$ & $\xi_A^d$ & $\xi_A^\ell$ \\ \hline
Type-I
& $c_\alpha/s_\beta$ & $+c_\alpha/s_\beta$ & $+c_\alpha/s_\beta$
& $s_\alpha/s_\beta$ & $s_\alpha/s_\beta$ & $s_\alpha/s_\beta$
& $-\cot\beta$ & $+\cot\beta$ & $+\cot\beta$ \\
Type-II
& $c_\alpha/s_\beta$ & $-s_\alpha/c_\beta$ & $-s_\alpha/c_\beta$
& $s_\alpha/s_\beta$ & $c_\alpha/c_\beta$ & $c_\alpha/c_\beta$
& $-\cot\beta$ & $-\tan\beta$ & $-\tan\beta$ \\
Type-X
& $c_\alpha/s_\beta$ & $+c_\alpha/s_\beta$ & $-s_\alpha/c_\beta$
& $s_\alpha/s_\beta$ & $s_\alpha/s_\beta$ & $c_\alpha/c_\beta$
& $-\cot\beta$ & $+\cot\beta$ & $-\tan\beta$ \\
Type-Y
& $c_\alpha/s_\beta$ & $-s_\alpha/c_\beta$ & $+c_\alpha/s_\beta$
& $s_\alpha/s_\beta$ & $c_\alpha/c_\beta$ & $s_\alpha/s_\beta$
& $-\cot\beta$ & $-\tan\beta$ & $+\cot\beta$ \\
\hline
\end{tabular}
\caption{\small Yukawa  couplings in the four types of 2HDM.} \label{Tab:MixFactor}
\end{table}

In order to avoid tree-level FCNC, it is natural to introduce a discrete $Z_2$ symmetry~\cite{Ref:GW}. All the possible nontrivial $Z_2$ charge assignments are listed in table~\ref{Tab:type}, which define the four well-known types of 2HDM, i.e. type-I, type-II, type-X and type-Y. The Yukawa interactions in the four models are different. In the mass-eigenstate basis, they can be unified in the form
\begin{align}
-{\mathcal L}_Y=&+\sum_{f=u,d,\ell} \left[m_f \bar f f+\left(\frac{m_f}{v}\xi_h^f \bar f fh+\frac{m_f}{v}\xi_H^f \bar f fH-i\frac{m_f}{v}\xi_A^f \bar f \gamma_5fA \right) \right]\nonumber\\
&+\frac{\sqrt 2}{v}\bar u \left (m_u V \xi_A^u P_L+ V m_d\xi_A^d P_R \right )d H^+ +\frac{\sqrt2m_\ell\xi_A^\ell}{v}\bar\nu_L \ell_R H^+
+\text{H.c.},\label{Eq:Yukawa}
\end{align}
where $P_{L,R}=(1\pm \gamma_5)/2$ and $V$ denotes the Cabibbo-Kobayashi-Maskawa (CKM) matrix. The couplings $\xi_{h,H,A}^f$ in the four types of 2HDM are listed in table~\ref{Tab:MixFactor}.

\section{Theoretical formalism for flavour observables}
\label{sec:flavourobservable}
In this section, we shall recapitulate the basic theoretical formulae for the relevant B-meson decay and mixing processes and discuss the  contributions of the four types of 2HDMs.
\subsection{$B_s-\bar B_s$ mixing}
\begin{figure}[t]
\centering
\includegraphics[width=0.24\textwidth]{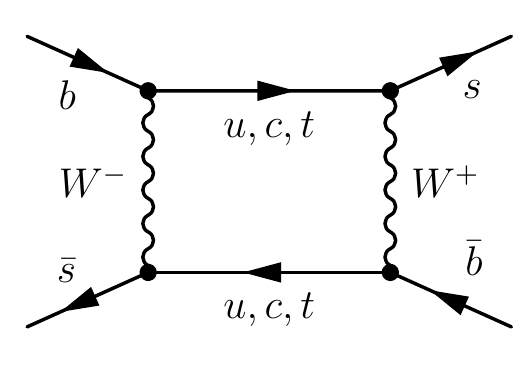}
\includegraphics[width=0.24\textwidth]{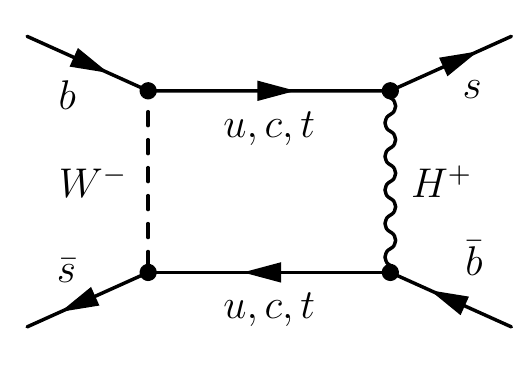}
\includegraphics[width=0.24\textwidth]{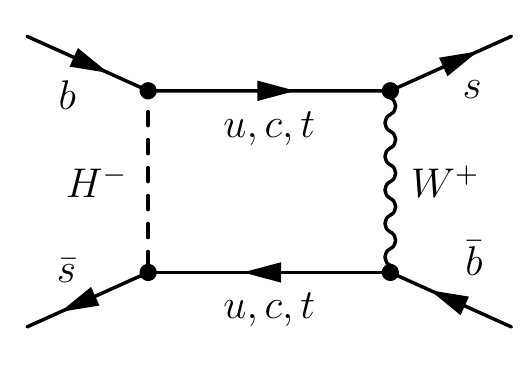}
\includegraphics[width=0.24\textwidth]{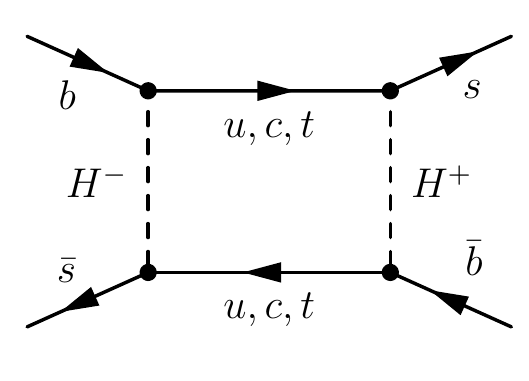}
\caption{\small Box diagrams for the $B_s-\bar B_s$ mixing in the SM and 2HDM.}
\label{LOboxen}
\end{figure}

For the $B_{s}-{\bar{B}}_{s}$ mixing, the mass difference is defined as
\begin{align}
\Delta m_{B_{s}}=m_{H}-m_{L},\label{Eq:deltmbs}
\end{align}
where $H$ and $L$ denote the heavy and light mass eigenstates. This quantity arises from $W$ box diagrams in the SM and can receive contributions from Higgs box diagrams in 2HDM, as shown in figure~\ref{LOboxen}. The theoretical prediction can be expressed as~\cite{Lenz:2010gu,Lenz:2012az,Urban:1997gw}
\begin{align}
\Delta m_{B_{s}}=\frac{G_F^2}{6\pi^2}m_W^2|V_{tb} V_{ts}^*|^2S(x_t, x_{H^{\pm}})\hat{\eta}_{B_s}\mathcal{B}_{B_{s}}(m_b) f^2_{B_s}m_{B_s},\label{Eq:m21}
\end{align}
with the definitions $x_t \equiv ({{\overline{m}}_t({\overline m}_{t})})^2/m_{W}^{2}$ and $x_{H^{\pm}} \equiv {{m}}_{H^{\pm}}^2/m_{W}^{2}$. The long-distance QCD effects are contained in the bag factor $\mathcal{B}_{B_{s}}(m_{b})$ and the decay constant $f_{B_s}$~\cite{Lenz:2010gu}. The short-distance contributions from the SM and 2HDM are encoded in the Inami-Lim function $S(x_t, x_{H^{\pm}})$, with its explicit expression given in appendix~\ref{appendix:1}, and the QCD correction factor $\hat{\eta}_{B_s}$.
\subsection{$\bar B\to X_s\gamma$ decay}
The effective Hamiltonian for $\bar{B}\rightarrow X_{s}\gamma$ at the scale $\mu_b={\mathcal O}(m_b)$ is given as follows~\cite{Li:2011fza,Deschamps:2009rh,Haisch:2002zz,Buras:2011we,Gambino:2001ew,Ciuchini:1997xe,Misiak:06}
\begin{figure}[t]
\centering
\includegraphics[width=0.24\textwidth]{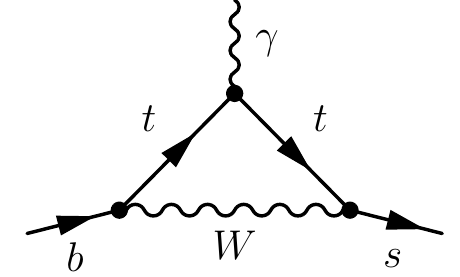}
\includegraphics[width=0.24\textwidth]{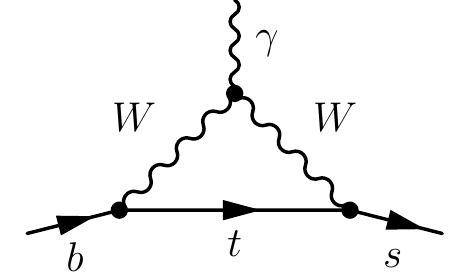}
\includegraphics[width=0.24\textwidth]{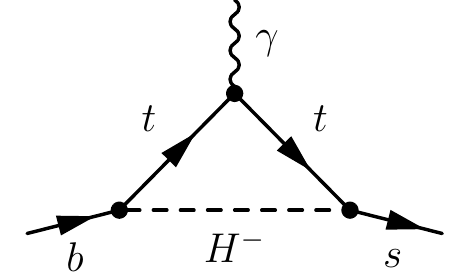}
\includegraphics[width=0.24\textwidth]{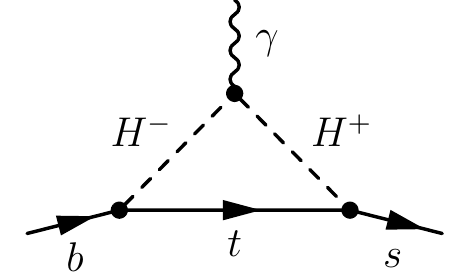}
\caption{\small One-loop diagrams contributing to $\bar B\to X_s\gamma$ through the $W$ boson and the charged Higgs boson exchange in the SM and 2HDM, respectively.}
\label{btosgamma}
\end{figure}
\begin{eqnarray}
\mathcal H_{\rm eff}=-\frac{G_F}{\sqrt{2}}V_{ts}^{\ast}V_{tb}\left(\sum\limits_{i=1}^{6} C_i (\mu_b)Q_i+C_{7\gamma} (\mu_b)
Q_{7\gamma} + C_{8g}(\mu_b)Q_{8g}\right),
\end{eqnarray}
where $Q_{1-6}$ are the four-fermion operators whose explicit expressions are given in ref.~\cite{Buras:2011we}. The remaining magnetic-penguin operators, which are characteristic for this decay, are defined as
\begin{eqnarray}
Q_{7\gamma}=\frac{e}{8\pi^2} m_b \bar{s}_{\alpha}\sigma^{\mu\nu}(1+\gamma_5)b_{\alpha}F_{\mu\nu}, \qquad Q_{8g} =\frac{g_s}{8\pi^2} m_b \bar{s}_{\alpha}\sigma^{\mu\nu}(1+\gamma_5)T^{a}_{\alpha\beta}b_{\beta}G_{\mu\nu}^{a},
\end{eqnarray}
where $m_b$ denotes the $b$-quark mass in the $\rm \overline {MS}$ scheme, and $e$ ($g_s$) is the electromagnetic (strong) coupling constant. The Wilson coefficients $\lbrace C_i \rbrace$ can be calculated perturbatively. In 2HDM, the photon-penguin diagrams mediated by charged Higgs, as shown in figure~\ref{btosgamma}, result in the following derivations:
\begin{eqnarray}
C_{7\gamma}=C_{7\gamma}^{\text{SM}}+C_{7\gamma}^{\text{2HDM}}, \qquad C_{8g}=C_{8g}^{\text{SM}}+C_{8g}^{\text{2HDM}}.
\end{eqnarray}
In the SM and the four types of 2HDM, analytic expressions for the Wilson coefficients up to the next-to-leading order (NLO) are given in refs.~\cite{Gambino:2001ew, Ciuchini:1997xe}. The next-to-next-leading order (NNLO) SM and 2HDM calculation can be found in ref.~\cite{Misiak:04} and ref.~\cite{2HDM:ph:6}, respectively.

The branching ratio of $\bar{B}\rightarrow X_{s}\gamma$ with an energy cut-off $E_0$ can be expressed as
\begin{eqnarray}
{\mathcal B}(\bar{B}\rightarrow X_{s} \gamma)_{{E_{\gamma}}\geq E_{0}} = {\mathcal B}(\bar{B}\rightarrow X_{c} e \bar{\nu})_{\rm exp}
{{\bigg \lvert}\frac{V_{ts}^{*} V_{tb}}{V_{cb}}{\bigg \rvert}}^2 \frac{6\alpha_e}{\pi C} [P(E_{0})+N(E_{0})],\label{Eq:bsgbr}
\end{eqnarray}
with the semi-leptonic factor
\begin{eqnarray}
C={\left\lvert\frac{V_{ub}}{V_{cb}} \right\rvert}^2
\frac{\Gamma(\bar B \to X_c e \bar \nu)}{\Gamma(\bar B \to X_u e \bar\nu)}.
\end{eqnarray}
The perturbative quantity $P(E_{0})$, which is expressed in terms of Wilson coefficients, and the non-perturbative correction $N(E_{0})$ can be found in ref.~\cite{Gambino:2001ew}.

\subsection{$B \to \tau\nu$ decay}
\begin{figure}[t]
\centering
\hspace{0cm}\includegraphics[width=0.24\textwidth]{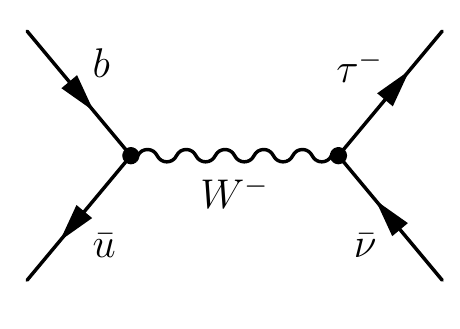}
\hspace{2cm}\includegraphics[width=0.24\textwidth]{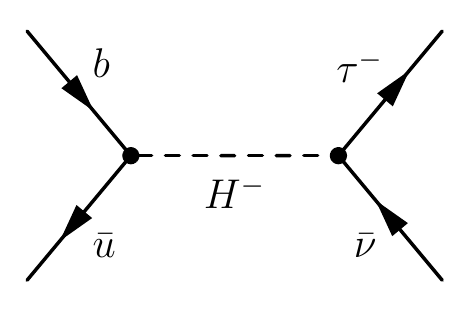}
\caption{\small Tree-level diagrams contributing to $B \to \tau \nu_{\tau}$ in the SM and 2HDM.}
\label{btotaunv}
\end{figure}

The tauonic decay $B \to \tau \nu$ is described as annihilation
processes mediated by $W$ boson in the SM and the charged Higgs boson
in 2HDM, as shown in figure~\ref{btotaunv}. Therefore, this process is
very sensitive to the charged Higgs boson $H^{\pm}$ and provides an important constraint on the model parameters.

Within 2HDM, the decay width of this channel reads~\cite{Ref:mayumiaoki,Akeroyd:2007eh,Akeroyd:2008ac},
\begin{align}
\Gamma ( B\to \tau \nu )=\frac{G_F^2{{\left| V_{ub} \right|}^{2}}}{8\pi }f_B^2m_Bm_\tau^2 {{\left( 1-\frac{m_\tau^2}{m_B^2} \right)}^2}{{\left( 1-\frac{m_B^2}{m_{H^\pm }^2} \xi_A^d\xi_A^\ell \right)}^2},\label{Eq:btaunu}
\end{align}
where $V_{ub}$ is the CKM matrix element and $f_B$ denotes the B-meson decay constant.
\subsection{$B_{s,d}\to \mu^+ \mu^-$ decay}
\begin{figure}[t]
\centering
\includegraphics[width=0.24\textwidth]{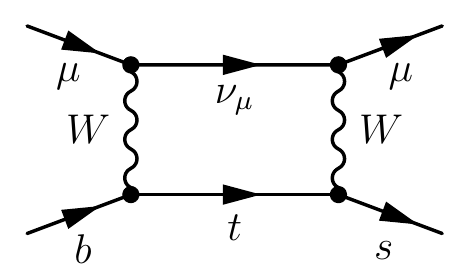}
\includegraphics[width=0.24\textwidth]{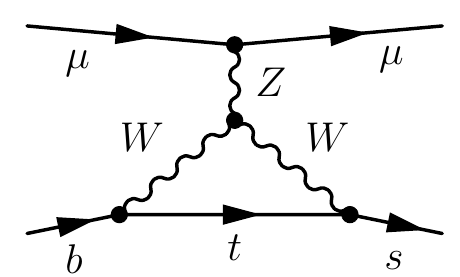}
\includegraphics[width=0.24\textwidth]{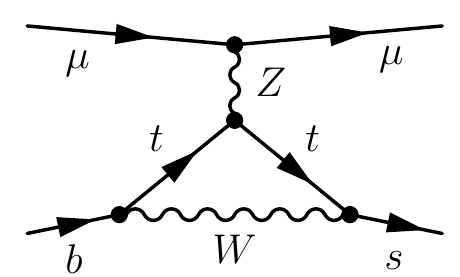}\\
\vskip 0.6cm
\includegraphics[width=0.24\textwidth]{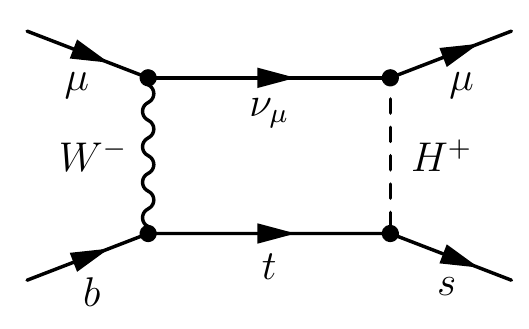}
\includegraphics[width=0.24\textwidth]{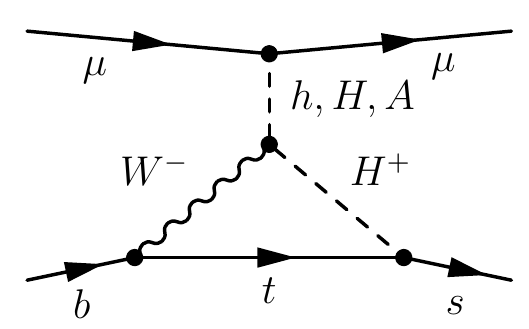}
\includegraphics[width=0.24\textwidth]{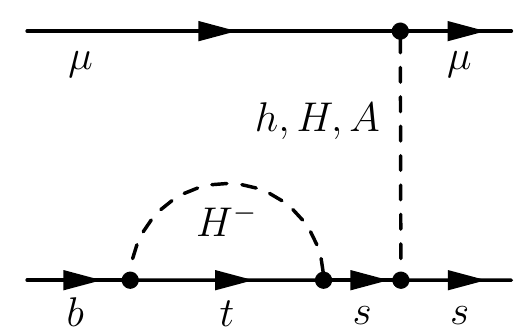}
\caption{\small Dominant SM and 2HDM diagrams for the $B_s\to \mu^+ \mu^-$ decays.}
\label{btomumusm}
\end{figure}

In the SM, the $B_q \to \mu^+ \mu^-$ decays ($q=d \text{ or } s$) arise from the $W$ box and $Z$ penguin diagrams at the quark level~\cite{Bobeth:2001sq,Buras:2012ru}, as shown in figure~\ref{btomumusm}. The helicity suppression in these decays may be relaxed by NP contributions, which can significantly enhance their branching ratios. Generally, the low-energy effective Hamiltonian for $B_q \to \mu^+ \mu^-$ decay can be written as ~\cite{Logan:2000iv}
\begin{align}
\mathcal H_{\rm eff}=\frac{G_F}{\sqrt{2}}\frac{\alpha_e}{2\pi s_W}V_{tb}^{\ast}V_{tq}\left( C_{S}Q_{S}+C_{P}Q_{P}+C_{A}Q_{A}\right), \label{Eq:bsmumueffham}
\end{align}
with $s_W \equiv \sin\theta_W$. The semi-leptonic operators are defined as
\begin{align}
Q_{S}=m_{b} (\bar{b}P_{L}q) (\bar{\mu}\mu), \qquad
Q_{P}=m_{b} (\bar{b}P_{L}q) (\bar{\mu}\gamma_{5}\mu), \qquad
Q_{A}=(\bar{b}\gamma^{\mu}P_{L}q) (\bar{\mu}\gamma_{\mu}\gamma_{5}\mu).\label{Eq:bsmumuoperator}
\end{align}
Among the Wilson coefficients $C_{S,P,A}$, only $C_A$ is non-zero in the SM. Its explicit expressions up to NLO can be found in refs.~\cite{Buchalla:1993bv,Misiak:99,Buras:99}. Recently, the NLO EW~\cite{Bobeth:13} and NNLO QCD~\cite{Misiak:13:2} corrections have also been completed~\cite{Misiak:13:1}. In the 2HDM, $C_A$ is not affected, whereas $C_{S,P}$ receive contributions from both charged and neutral Higgs bosons. At present, only the diagrams shown in figure~\ref{btomumusm} have been calculated in the type-II 2HDM with large $\tan\beta$~\cite{Logan:2000iv}. Based on these results, we give the Wilson coefficients $C_{S,P}$ corresponding to these diagrams in all the four types of 2HDM with arbitrary $\tan\beta$ in appendix~\ref{appendix:2}. It is noted that contributions from other diagrams  may be important for some specific values of $\tan\beta$ (large or small) and will become crucial with future high-precision measurement of $B_q \to \mu^+ \mu^-$ decays.

For $B_q \to \mu^+ \mu^-$ decays, one important observable is the CP averaging branching ratio, which reads
\begin{align}
\mathcal B (B_q\to \mu^+ \mu^-)=\frac{G_F^2 {\alpha_e^2}}{32\pi^2 s_W^4}
\frac{m_{B_{q}}^3 \tau_{B_{q}} f_{B_{q}}^2}{8\pi} \sqrt{1-\frac{4 m_{\mu}^2}{m_{B_{q}}^2}} \left(\frac{2m_\mu}{m_{B_q}} \right)^2 \abs{V_{tb} V_{tq}^*}^2 \abs{C_A}^2 (\abs{S}^2+ \abs{ P }^2),
\end{align}
with the definitions
\begin{align}
P \equiv 1-\frac{m_{B_q}^2}{2m_{\mu}}\frac{C_P^{\ast}}{C_A^{\ast}}, \qquad
S \equiv\sqrt{1-4\frac{m_{\mu}^2}{m_{B_q}^2}}\frac{m_{B_q}^2}{2m_{\mu}}\frac{C_S^{\ast}}{C_A^{\ast}}.
\end{align}
It is noted that the contributions of the $C_{S,P}$ terms do not
suffer helicity suppression, but they are suppressed by the small
leptonic Yukawa coupling in the 2HDMs. However, they may be enhanced
by a large $\tan\beta$ (or $\cot\beta$) factor~\cite{Bobeth:2001sq,Buras:2012ru,Buchalla:1993bv,Misiak:99,Logan:2000iv}.

Recently, a sizable width difference $\Delta \Gamma_s$ between the $B_s$ mass eigenstates has been measured at the LHCb \cite{LHCb:13}
\begin{align}
  y_s \equiv \frac{\Gamma_s^{\rm L}- \Gamma_s^{\rm H}}{\Gamma_s^{\rm L}+ \Gamma_s^{\rm H}}=\frac{\Delta\Gamma_s}{2\Gamma_s}=0.080 \pm 0.010,
\end{align}
where $\Gamma_s$ denotes the inverse of the $B_s$ mean lifetime $\tau_{B_s}$. As pointed out in ref.~\cite{DeBruyn:2012wk}, the measured branching ratio of $B_q \to \mu^+ \mu^-$ should be the time-integrated one, denoted by $\overline { \mathcal B} (B_q \to \mu^+\mu^-)$. For $B_s \to \mu^+ \mu^-$ decay, in order to compare with the experimental measurement, the sizable width difference effect should be taken into account in the theoretical prediction, and one has
\begin{align}
&\overline{ {\mathcal B}}({{{B}}_{s}}\to \mu^+\mu^-)=\left[\frac{1+A_{\Delta\Gamma}y_{s}}{1-y_{s}^2} \right] {\mathcal B}({{{B}}_{s}}\to \mu^+\mu^-),\label{Eq:bstomumubmix}
\end{align}
where $A_{\Delta\Gamma}$ denotes the mass eigenstate rate asymmetry,
which can be expressed as
\begin{align}\label{eq:adeltgamma}
A_{\Delta\Gamma}=\frac{|P|^2\cos 2\varphi_P-|S|^2\cos 2\varphi_S}{|P|^2+|S|^2},
\end{align}
where $\varphi_P$ and $\varphi_S$ denote the phase of the quantity $P$
and $S$, respectively. In the four types of 2HDM,
$\varphi_P=\varphi_S=0$. The observable $A_{\Delta\Gamma}$ is
complementary to the branching ratio of $B_s \to \mu^+ \mu^-$,
offering independent information on the short-distance structure of
this decay. It can be extracted from the time-dependent untagged decay
rate~\cite{DeBruyn:2012wk, Buras:2013uqa}. In the SM,
$A_{\Delta\Gamma} = +1$. In addition, since the finite width
difference in the $B_d$ system is negligible, the approximation $\overline{\mathcal B}(B_d\to \mu^+\mu^-)\approx{\mathcal B}(B_d\to \mu^+\mu^-)$ works well.

Following ref.~\cite{DeBruyn:2012wk}, it is convenient to introduce the ratio
\begin{align}\label{eq:ratioofexpandsm}
R \equiv \frac{\overline{{\mathcal B}}(B_s\rightarrow \mu^+\mu^-)}{{\mathcal B}(B_s\rightarrow \mu^+\mu^-)_{\text{SM}}}=\frac{1+y_s \cos 2\varphi_P}{1-y_s^2}|P|^2+\frac{1-y_s \cos 2\varphi_S}{1-y_s^2}|S|^2,
\end{align}
where $\varphi_P=\varphi_S=0$ in the four types of 2HDM.

It is also useful to define the following quantity:
\begin{align}
R_{sd}\equiv\frac{\overline{{\mathcal B}}(B_s \to \mu^+\mu^-)}{\overline{{\mathcal B}}(B_d\rightarrow \mu^+\mu^-)} ,\label{Eq:ratiodefine}
\end{align}
in which some uncertainties of input parameters are canceled out. For example, the $f_{B_s} / f_{B_d}$ in the above ratio can be directly determined by Lattice QCD and the corresponding theoretical uncertainty is significantly reduced \cite{LQCD:1,LQCD:2}.
\section{Numerical analysis and discussions}
\label{sec:numericalanalysis}
With the theoretical framework presented in the previous sections, we proceed to present our numerical results and discussion in this section.
\begin{table}[deltbs]
\centering
\begin{tabular}{|lll|lll|}\hline
$V_{us}$                                      &$0.22537 \pm 0.00063$                    &\cite{Ref:utfit}       &
$s_W^2$                           & $0.23116\pm 0.00012$                      &\cite{Beringer:1900zz}
\\
$V_{ub}$                                 &$(0.00399 \pm 0.00055)e^{i(-71.1\pm5.1)^\circ}$   &\cite{Ref:utfit}       &
$\alpha_s (m_Z)$                                  & $0.1184\pm 0.0007$                  &\cite{Beringer:1900zz}
\\
$V_{cb}$                                      & $0.04071\pm 0.00096$                    &\cite{Ref:utfit}       &
$\alpha_e (m_Z)^{-1}$                          & $127.944\pm0.014$                   &\cite{Beringer:1900zz}
\\
$V_{td}$                                 & $(0.00872 \pm 0.00041)e^{i(-24.6\pm 2.7)^\circ}$ &\cite{Ref:utfit}       &
$f_{B_{s}}$                                 & $(227.6    \pm 5.0     ) \,\text{MeV}$         &\cite{Laiho:2009eu}
   \\
$V_{ts}$                                 & $(-0.03998 \pm 0.00094)e^{i(1.19\pm0.11)^\circ}$ &\cite{Ref:utfit}     &
$f_{B_d}$                                   & $(190.6\pm 4.7) \,\text{MeV}$                &\cite{Laiho:2009eu}
   \\
$V_{tb}$                                 & $0.999163  \pm 0.000039$                     &\cite{Ref:utfit}       &
$f_{B_s}/f_{B_d}$                           & $1.201\pm 0.017 $                         &\cite{Laiho:2009eu}
\\
$\overline{m}_{s}(\overline{m}_{b})$              & $(0.085 \pm 0.017) \,\text{GeV}$       &\cite{Lenz:2010gu}&
${\mathcal B}_{B_s}(m_{b})$             & $0.841   \pm 0.024$       &\cite{Lenz:2010gu}
\\
${\overline m}_{c}({\overline m}_{c})$ & $(1.275 \pm 0.025) \,\text{GeV}$       &\cite{Beringer:1900zz}&
$\hat{\eta}_{B}$                                  & $0.8393\pm 0.0034$               &\cite{Lenz:2010gu}
\\
$\overline{m}_{b}(\overline{m}_{b})$              & $(4.248 \pm 0.051) \,\text{GeV}$       &\cite{Lenz:2010gu} &
$\mathcal B (\bar{B}\to X_c e \bar\nu)$ & $0.101\pm0.004$                         &\cite{Beringer:1900zz}
\\
$m_{t}^{\text{pole}}$                                    & $(173.5 \pm 0.6 \pm 0.8) \,\text{GeV}$ &\cite{Beringer:1900zz} &
$m_{b}^{\rm 1S}$                                      & $(4.65 \pm 0.03) \,\text{GeV}$         &\cite{Beringer:1900zz}
\\
\hline
\end{tabular}
\caption{\small The relevant input parameters used in the numerical analysis. The meson masses and lifetimes can be found in ref.~\cite{Beringer:1900zz}.}
\label{tab:TheoreticalInputsofdeltmbs}
\end{table}
\subsection{SM predictions and experimental data}
\label{subsection:SM and experiment}
\subsubsection{Flavour observables within the SM}
Within the SM, our predictions for the flavour observables as well as the corresponding experimental data are collected in table~\ref{tab:smandexperiment}. The theoretical uncertainties are obtained by varying the input parameters listed in table~\ref{tab:TheoreticalInputsofdeltmbs} within their respective ranges and adding them in quadrature. It is noted that, taking into account the theoretical uncertainties, our SM predictions are in good agreement with the current data. The only tension appears in the branching ratio of $B_d\to \mu^+\mu^-$, which however has a rather large experimental error. Thus, strong constraints on the four types of 2HDM and good discrimination between them are excepted.
\begin{table}[t]
\centering
\begin{tabular}{|l l l c|}\hline
Observable                      &SM prediction              & Experiment                  & Ref.                             \\
\hline
$\Delta m_{B_s}$ [$10^{-11} \, \rm GeV$]    & $1.100_{-0.077}^{+0.079}$  &$1.164\pm 0.005$   &\cite{Beringer:1900zz}     \\
${\mathcal B}(B \to \tau \nu_\tau)$ [$10^{-4}$] & $1.02_{-0.27}^{+0.31}$  & $1.65 \pm 0.34$                        &\cite{Beringer:1900zz}     \\
${\mathcal B}(\bar{B}\to X_s \gamma)$  [$10^{-4}$]  &$3.16\pm0.26$        & $3.43\pm 0.22$                     &\cite{Ref:HFAGbsg}     \\
$\overline{{\mathcal B}}(B_d \to \mu^+ \mu^-)$ [$10^{-10}$]& $1.16_{-0.12}^{+0.13}$  &$3.6_{-1.4}^{+1.6}$              &\cite{CMSandLHCbCollaborations:2013pla,Aaij:2013aka,Chatrchyan:2013bka}  \\
$\overline{{\mathcal B}}(B_s \to \mu^+ \mu^-)$ [$10^{-9}$] &$3.76_{-0.25}^{+0.26}$ &$2.9\pm 0.7$               &\cite{CMSandLHCbCollaborations:2013pla,Aaij:2013aka,Chatrchyan:2013bka}     \\
$R_{sd}$&$32.84_{-3.81}^{+3.45}$&  &   \\
$R$&$1.08\pm0.01$&$0.86\pm 0.21$ & \cite{Aaij:2013aka,Chatrchyan:2013bka}  \\
\hline
\end{tabular}
\caption{\small SM predictions and experimental data for the flavour observables. For the inclusive $\bar B \to X_s \gamma$ decay, the value given here corresponds to a photon energy cut at $E_0=1.6 \,\rm GeV$.}
\label{tab:smandexperiment}
\end{table}
\subsubsection{Direct search for the Higgs bosons}
\label{directsearch}
Direct searches for charged Higgs bosons motivated by 2HDM have been
performed at LEP~\cite{Heister:2002ev},
Tevatron~\cite{Abulencia:2005jd,Abazov:2009aa} and
LHC~\cite{TheATLAScollaboration:2013wia,Chatrchyan:2012vca}. However,
the obtained limits on the charged-Higgs mass depend strongly on the
assumed Yukawa structure. In type-II 2HDM, the parameter space with
$m_{H^\pm}<m_t$ is almost excluded by the
ATLAS~\cite{TheATLAScollaboration:2013wia}, which, however, can not be
readily translated into constraints on the parameters of other
2HDMs. Without assumptions on the Yukawa structure, the LEP collaboration established the bound on the charged Higgs boson mass~\cite{Heister:2002ev}
\begin{align*}
m_{H^\pm} \geq79.3\,\text{GeV},
\end{align*}
in which $\mathcal B(H^+\to\tau^+\nu_\tau)+\mathcal B(H^+\to c\bar s)=1$ is assumed. In addition, the hadronic $Z\to b\bar b$ branching ratio $R_b$ can also set indirect limits on $m_{H^\pm}$. However, the bounds from $R_b$ are weaker than that from the $B_s-\bar B_s$ mixing~\cite{Deschamps:2009rh}.

Recently, the LHC and Tevatron data collected so far~\cite{ATLAS:1,ATLAS:2,CMS:1,CMS:2,Aaltonen:2012qt} confirm the SM Higgs-like nature \cite{Higgs:64:1,Higgs:64:2,Higgs:66,Englert:1964et,Guralnik:1964eu,Kibble:1967sv} of the new boson discovered at the LHC, with a spin/parity consistent with the SM $0^+$ assignment ~\cite{Aad:2013xqa,Chatrchyan:2012jja,D0-6387}. The observation of its $\gamma\gamma$ decay mode demonstrates that it is a boson with $J\not=1$, while the $J^P=0^-$ and $2^+$ hypotheses have been already excluded at about 99\% CL, by analyzing the distribution of its decay products. The masses measured by ATLAS and CMS are in good agreement, giving the average value~\cite{Pich:2013vta}
\begin{align*}
m_h = (125.64 \pm 0.35)\, \rm GeV.
\end{align*}
If the light neutral Higgs boson $h$ in 2HDM is identified as the observed resonance at LHC, the decoupling limit $\sin(\beta-\alpha)=1$ is needed to keep its Yukawa couplings SM-like \cite{Zupan:12,Craig:13,2HDM:ph:14}.
\subsubsection{Perturbative unitarity and vacuum stability}
Besides the experimental constraints mentioned in previous sections, there are theoretical conditions which allow one to restrict the 2HDM parameter space~\cite{Ref:mayumiaoki,Chang:2012ve,Kanemura:1999xf,Ref:THDMreport}. The vacuum stability~\cite{Sher:1988mj} arises from the requirement that the Higgs potential must have a minimum. The perturbative unitarity~\cite{Ref:LQT} is the condition that all the (tree-level) scalar-scalar scattering amplitudes must respect unitarity. From these conditions, the following bound can be obtained,
\begin{align*}
  |y_{t}|^2\leq 4\pi \qquad \text{or} \qquad \tan\beta \geq 0.28,
\end{align*}
with $|y_t| \equiv (\sqrt 2 \overline m_t(m_t^{\rm pole}))/(v \sin\beta)$~\cite{Kanemura:1999xf}.
\subsection{Procedure in numerical analysis}
\label{sec:procedure}
As shown in section~\ref{sec:formalism}, the relevant 2HDM parameters contain two angles, $\alpha$ and $\beta$, and four mass parameters $m_{H^\pm}$, $m_h$, $m_H$ and $m_A$, corresponding to the mass of charged Higgs $H^\pm$, light neutral Higgs $h$, heavy neutral Higgs $H$, and CP-odd neutral Higgs $A$. As discussed in ref.~\cite{Zupan:12,Craig:13}, we choose the light neutral Higgs boson $h$ as the observed resonance at LHC and take the decoupling limit $\sin(\beta-\alpha)=1$. Then the parameter space is reduced to ($m_H$, $m_A$, $m_{H^\pm}$, $\tan\beta$) and we shall restrict these parameters in the following ranges:
\begin{align*}
    m_H\in [m_h,1000]\,\text{GeV},\quad m_A\in [1,1000]\,\text{GeV},\quad m_{H^\pm}\in [1,1000]\,\text{GeV} ,\quad \tan\beta\in [0.1,100].
\end{align*}

In the numerical analysis, we impose the experimental constraints in
the following way: each point in the parameter space corresponds to a
theoretical range, constructed from the prediction for the observable
in that point together with the corresponding theoretical
uncertainty. If this range overlaps with the $2\sigma$ range of the
experimental measurement, this point is regarded as allowed. In this
procedure, to be conservative, the theoretical uncertainty is taken as
twice the one listed in table~\ref{tab:smandexperiment}. Since the
main theoretical uncertainties arise from hadronic inputs, common to
both the SM and the 2HDM, the relative theoretical uncertainty is assumed constant over the parameter space.
\subsection{$B_s-\bar B_s$ mixing within 2HDM}
\begin{figure}[t]
\centering
\includegraphics[width=0.43\textwidth]{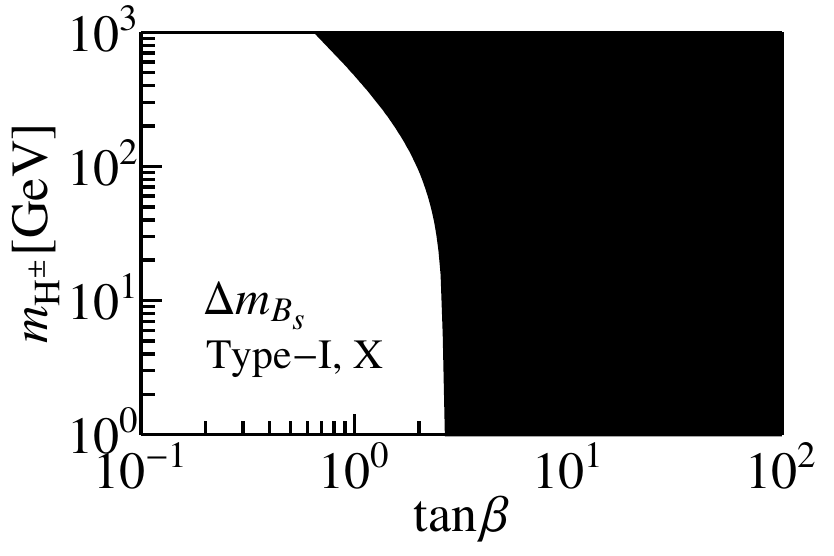}
\qquad
\includegraphics[width=0.43\textwidth]{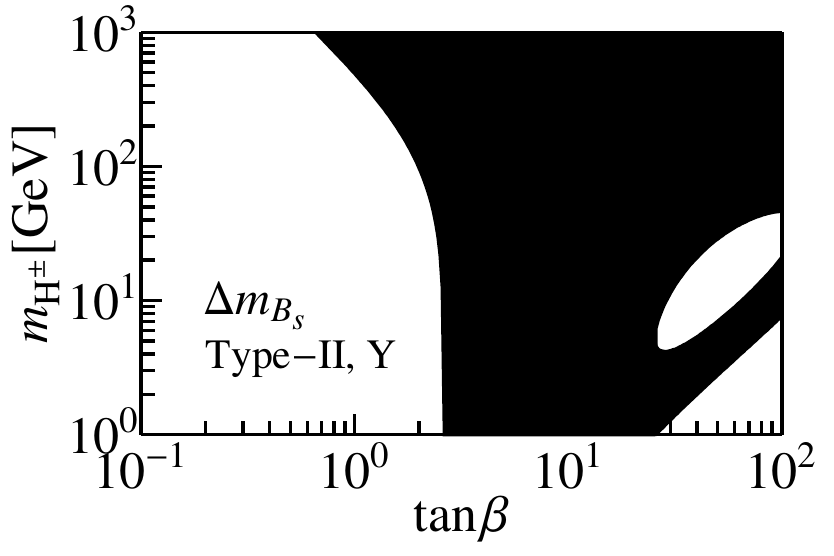}
\caption{\small Constraints on the parameter space $(\tan\beta,m_{H^\pm})$ of the four types of 2HDM from $\Delta m_{B_s}$. The allowed regions are shown in \textit{black}.}
\label{resultofbmixing}
\end{figure}
The mixing parameter $\Delta m_{B_{s}}$ is proportional to the Inami-Lim function $S(x_t,x_{H^\pm})$. In the leading order (LO) approximation and taking $m_{H^{\pm}}=500\,\text{GeV}$, we have numerically
\begin{align*}
\frac{S(x_t, x_{H^\pm})}{S_{\rm SM}(x_t, x_{H^\pm})}=
\begin{cases}
\displaystyle 1+\frac{3.5\times 10^{-2}}{\tan^4\beta}+\frac{0.2}{\tan^2\beta},\hspace{0.1cm}&\text{type-I, X},\\
\displaystyle 1+\frac{3.5\times10^{-2}}{\tan^4\beta}+\frac{0.2}{\tan^2\beta}+1.6\times10^{-6}\tan^2\beta,\hspace{0.1cm}&\text{type-II, Y}.
\end{cases}
\end{align*}
From these results, we make the following observations:
\begin{itemize}
\item For the four different 2HDMs, the dominant effect is proportional to $\cot\beta$. They always work constructively with the SM contribution, even when the charged Higgs mass $m_{H^\pm}$ is not fixed but larger than about $90\, \rm GeV$.
\item Since $\Delta m_{B_s}$ is only affected by charged Higgs, the
  contributions from type-I and -X (type-II and -Y) 2HDMs are the
  same. The type-I and -X Yukawa couplings of down-type quarks are
  different from the type-II and -Y ones. Thus, there is an additional term proportional to $\tan\beta$ in the latter two 2HDMs, however, suffering from down-type quark mass suppression.
\end{itemize}

In figure~\ref{resultofbmixing}, the constraints on the parameter space $(\tan\beta,m_{H^\pm})$ from $\Delta m_{B_s}$ are shown. As expected, the allowed parameter space in type-I, -X 2HDMs and type-II, -Y 2HDMs are almost the same, in which the regions with small $\tan\beta$ are excluded. The difference appears in the region with large $\tan\beta$. However, the allowed charged Higgs mass in this region is below the LEP lower limit.
\subsection{$\bar B\to X_s \gamma$ decay within 2HDM}

The branching ratio of $\bar B\to X_s\gamma$ decay is proportional to  $|C_{7\gamma}^{\text{eff}}(\mu_b)|^2$ in the LO approximation. In 2HDM, the Wilson coefficient $C_{7\gamma}^{\text{eff}}(\mu_b)$ reads numerically at $m_{H^\pm}=500\,\text{GeV}$ in the LO,
\begin{align*}
\frac{C_{7\gamma}^{\text{eff}}(\mu_b)}{C_{7\gamma,\text{SM}}^{\text{eff} }(\mu_b)}=
\begin{cases}
\displaystyle 1+\frac{0.02}{\tan^2\beta}-\frac{0.18}{\tan^2\beta},&\text{type-I, X},\\
\displaystyle 1+\frac{0.02}{\tan^2\beta}+0.18,&\text{type-II, Y}.
\end{cases}
\end{align*}
From these numerical results, we make the following observations:
\begin{itemize}
\item In the type-I and -X models, the 2HDM effect is proportional to $\cot\beta$ and destructive with the SM contribution.
\item In the type-II and -Y models, the 2HDM contribution works constructively with the SM one. Besides the $\tan\beta$ terms, there are also $\beta$-independent terms, which dominate the 2HDM contribution for large $\tan\beta$.
\item For the $\bar B \to X_s \gamma$, unlike the case of $B_s -\bar B_s$ mixing, the dominant operator $Q_{7\gamma}$ is a chirality-flipped operator with the chirality transition $b_R \to s_L$. Thus the contributions from down-type quark Yukawa couplings do not suffer mass suppression and dominate the ones from up-type quark Yukawa couplings.
\end{itemize}
\begin{figure}[t]
\centering
\includegraphics[width=0.43\textwidth]{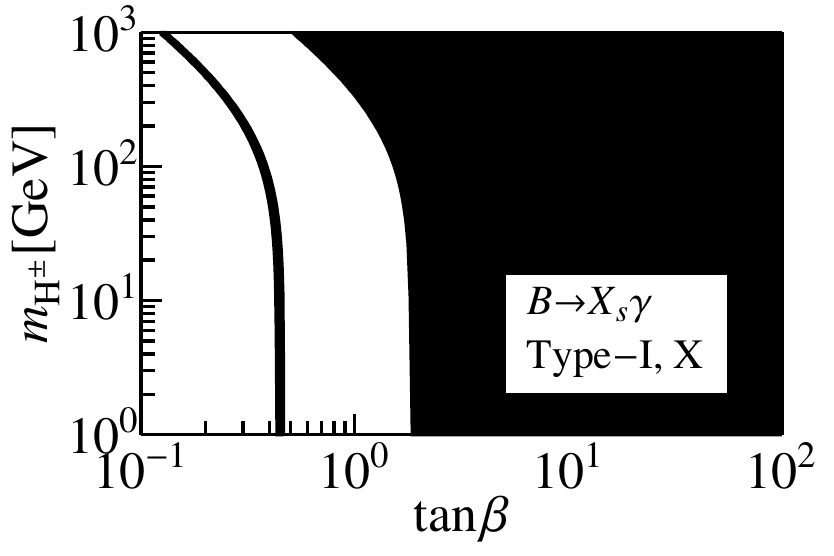}
\qquad
\includegraphics[width=0.43\textwidth]{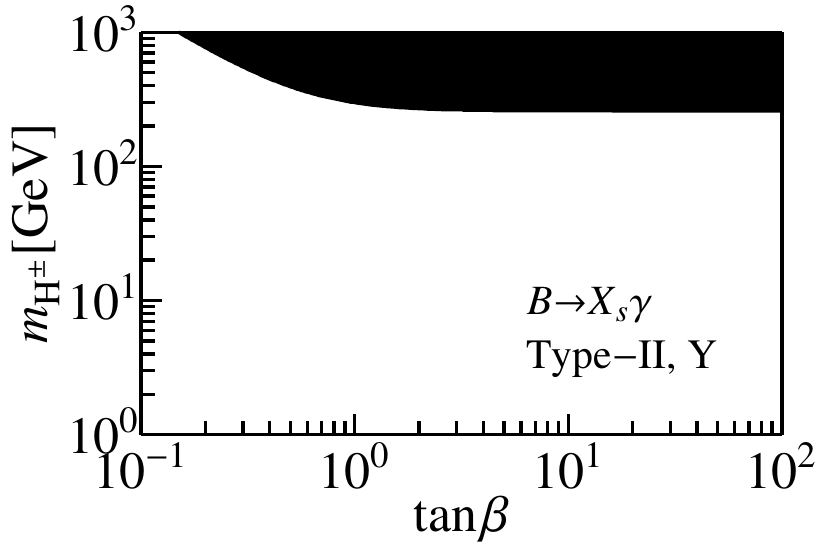}
\caption{\small Constraints on the parameter space $(\tan\beta,m_{H^\pm})$ of the four types of 2HDM from ${\mathcal B}(\bar B\to X_s \gamma)$. The allowed regions are shown in \textit{black}.}
\label{resultofbtosgamma}
\end{figure}
In figure~\ref{resultofbtosgamma}, the constraints on the parameter space $(\tan\beta, m_{H^\pm})$ from $\mathcal B(\bar B\to X_s\gamma)$ are shown. The regions with small $\tan\beta$ are largely excluded in all the four types. However, there is still one solution in the type-I and -X 2HDMs, where the destructive interference between the SM and 2HDM contributions makes the coefficient $C_{7\gamma}^{\rm eff}$ sign-flipped. For the type-II and -Y 2HDMs, the charged Higgs mass is strongly bounded,
\begin{align}
m_{H^\pm}\geq 259 \, \text{GeV},\nonumber
\end{align}
which mainly arises from the $\beta$-independent terms. This lower limit is stronger than the LEP bound.
%

\subsection{$B\to \tau\nu$ decay within 2HDM}


For $B\to \tau\nu$ decay, the numerical expressions of the branching ratio read,
\begin{align*}
\frac{\mathcal B ( B \to \tau \nu )}{\mathcal B ( B \to \tau \nu )_{\text{SM}}}=
\begin{cases}
\displaystyle (1-\frac{27.9}{m_{H^\pm}^{2}\tan^2\beta})^2=(1-\frac{1.1\times 10^{-4}}{\tan^2\beta})^2, &\text{type-I},\\
\displaystyle (1-\frac{27.9\tan^2\beta}{m_{H^\pm}^{2}})^2=(1-1.1\times 10^{-4}{\tan^2\beta})^2,&\text{type-II},\\
\displaystyle (1+\frac{27.9}{m_{H^\pm}^{2}})^2=(1+1.1\times 10^{-4})^2,&\text{type-X, Y},
\end{cases}
\end{align*}
in which the second equality in each line holds for $m_{H^\pm}=500 \, \text{GeV}$. Here, the 2HDM effects arise from tree-level charged Higgs with leptonic couplings, which make the following features:
\begin{itemize}
\item In all the four types, the 2HDM effects are largely suppressed by the charged Higgs mass.
\item In the type-II (-I) model, the 2HDM effect is constructive with the SM one and proportional to $\tan\beta$ ($\cot\beta$). The large (small) $\tan\beta$ can compensate the mass suppression.
\item In the type-X and -Y model, the 2HDM contribution is $\beta$-independent and proportional to $1/m_{H^\pm}$. Thus, small $m_{H^\pm}$ is expected to be strongly bounded.
\end{itemize}

\begin{figure}[t]
\centering
\includegraphics[width=0.32\textwidth]{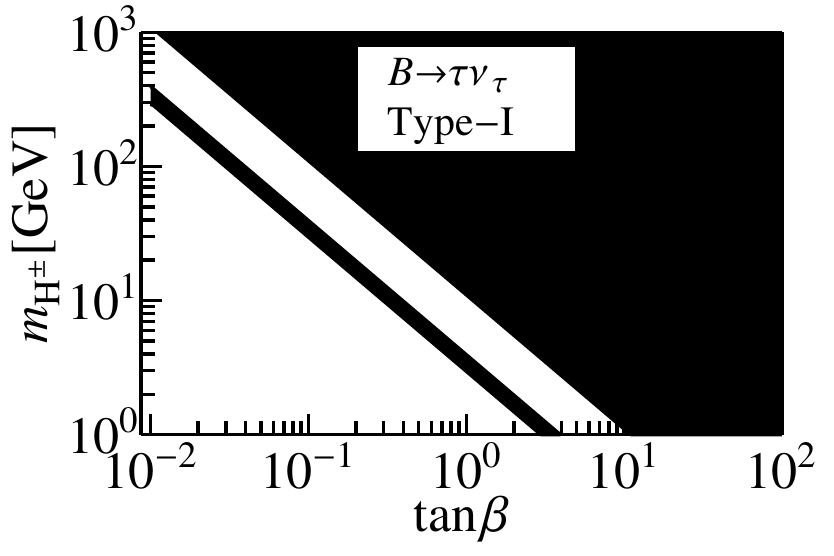}
\includegraphics[width=0.32\textwidth]{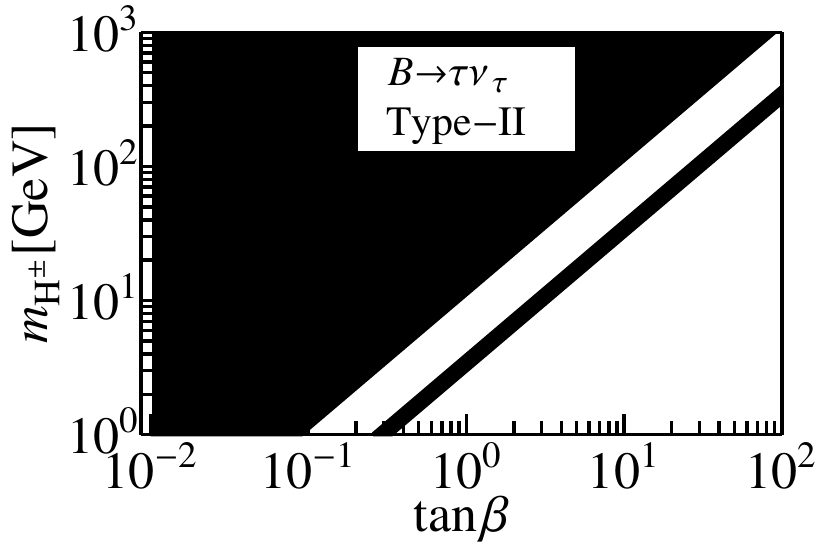}
\includegraphics[width=0.32\textwidth]{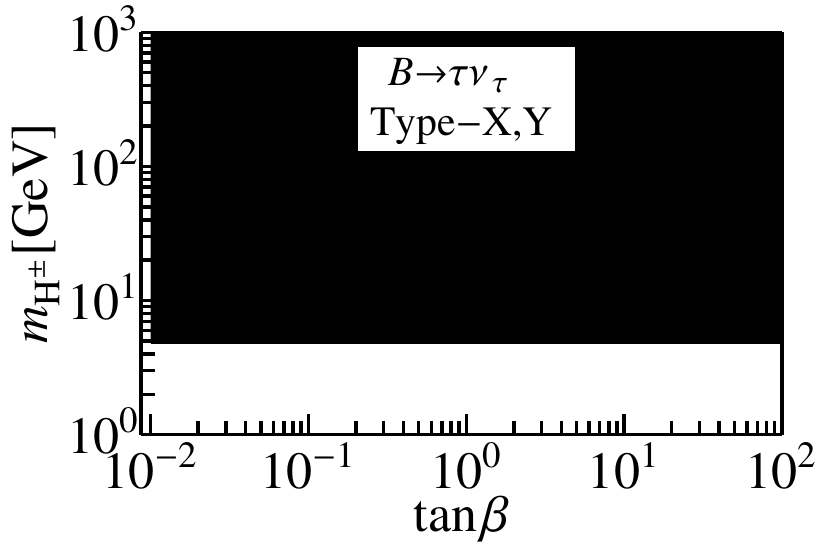}
\caption{\small Constraints on the parameter space $(\tan\beta,m_{H^\pm})$ of the four types of 2HDM from $\mathcal B( B \to \tau \nu)$. The allowed regions are shown in \textit{black}.}
\label{resultofbtotaunv}
\end{figure}

The constraints on the parameter space $(\tan\beta,m_{H^\pm})$ from
${\mathcal B}(B\to \tau\nu)$ are shown in
figure~\ref{resultofbtotaunv}\footnote{The bounds derived in this
  paper are weaker than the ones in the literature, for example in
  ref.~\cite{2HDM:ph:8}, since a conservative procedure is used in the
  numerical analysis, which is explained in detail in sec.~\ref{sec:procedure}.}. As expected, in the type-I (II) 2HDM, excluded regions mainly arise from the parameter space with small (large) $\tan\beta$. There also exists one solution (narrow band in figure~\ref{resultofbtotaunv}), where the sign of the SM contribution is flipped by the 2HDMs.  For the type-X and -Y 2HDMs, a $\beta$-independent bound on the charged Higgs mass is obtained, $m_{H^\pm}\geq 5\, \rm GeV$. However, this lower limit is much weaker than the LEP bound.
\begin{figure}[t]
\centering
\includegraphics[width=0.43\textwidth]{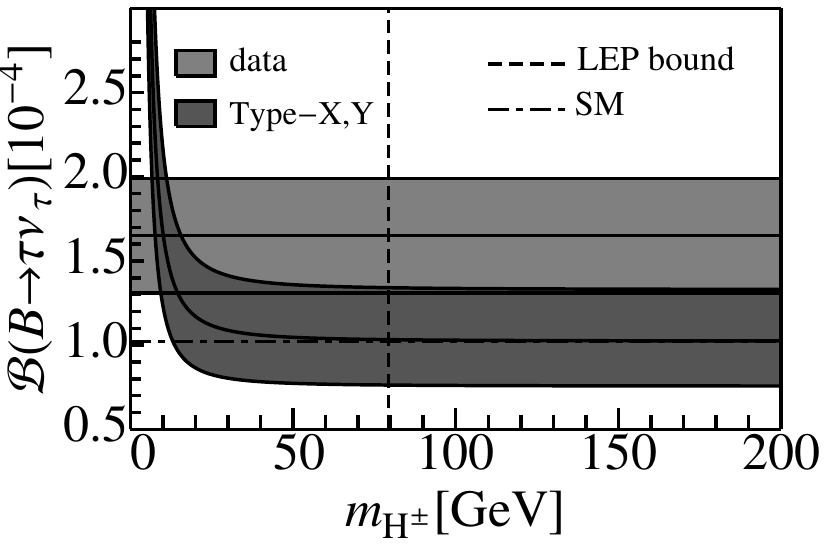}
\caption{\small The type-X and -Y 2HDM predictions on $\mathcal B (B\to \tau\nu)$ with the theoretical uncertainty (\textit{dark shaded band}) versus the experimental measurement (\textit{light shaded band}).}
\label{resultofbtotaunvxy}
\end{figure}

Since $\mathcal B(B\to \tau\nu)$ is independent of $\tan\beta$ in type-X and type-Y, we also present its theoretical prediction as a function of $m_{H^\pm}$ in figure~\ref{resultofbtotaunvxy}, which may be helpful for understanding these two models with reduced experimental and theoretical uncertainties in the future.


\subsection{$B_{s,d}\to \mu^+\mu^-$ decays within 2HDM}
\label{sec:bstomumu}

\begin{figure}[t]
\centering
\includegraphics[width=0.32\textwidth]{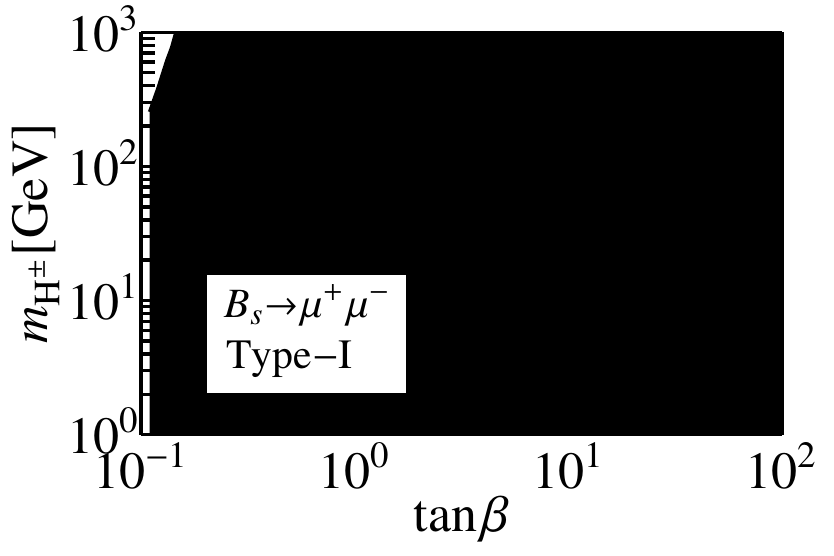}
\includegraphics[width=0.32\textwidth]{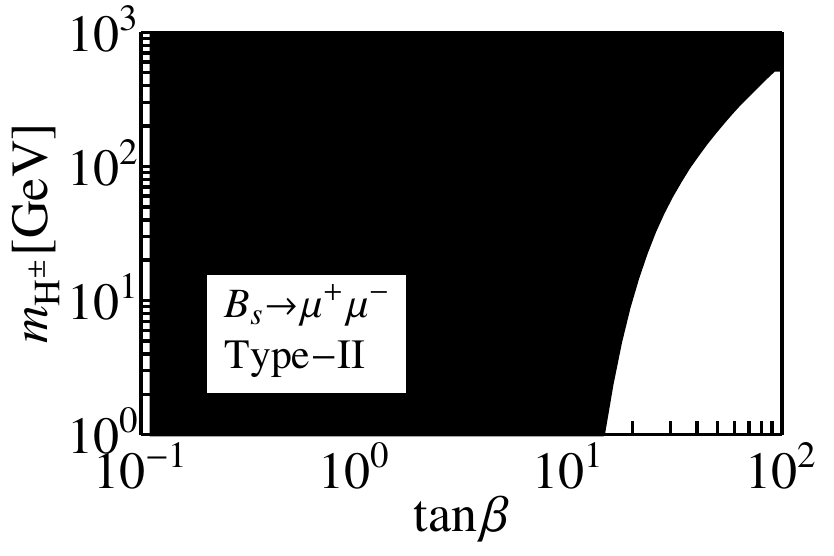}
\includegraphics[width=0.32\textwidth]{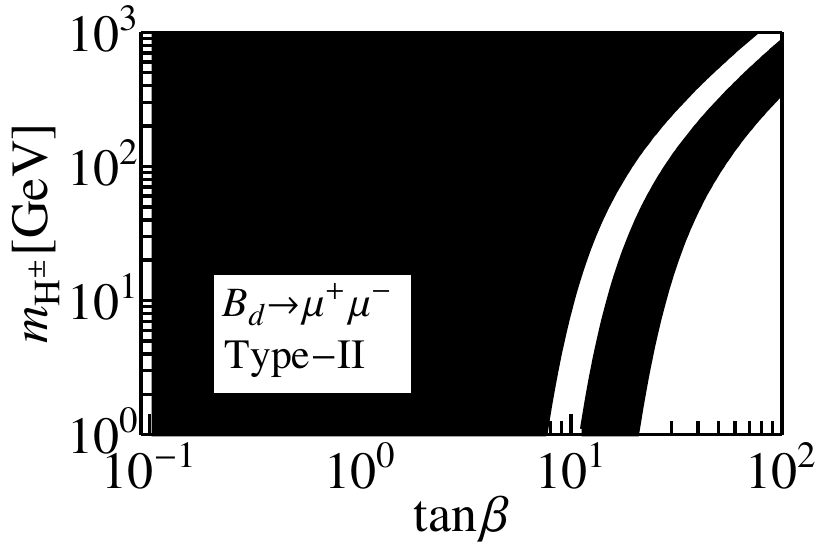}
\caption{\small Constraints on the parameter space of the four types of 2HDM from $\overline{\mathcal B}(B_{s,d}\to \mu^+\mu^-)$, plotted in the $\tan\beta-m_{H^\pm}$ plane. The allowed regions are shown in \textit{black}.}
\label{resultofbstomumu}
\end{figure}

For $B_s\to \mu^+\mu^-$ decay, taking $m_{H^\pm}=m_H=m_A=500\,\text{GeV}$, we get numerically
\begin{align*}
\frac{\overline{\mathcal B} (B_s \to \mu^+ \mu^-)}{\overline{\mathcal B}(B_s \to \mu^+ \mu^-)_{\rm SM}}=
\begin{cases}
\displaystyle 1-\frac{5.5\times10^{-5}}{\tan^2\beta}+\frac{2.5\times10^{-4}}{\tan^4\beta}+\frac{4.1\times10^{-7}}{\tan^6\beta}+\frac{2.9\times10^{-8}}{\tan^8\beta},&\text{type-I},\\
\displaystyle 1+3.4\times10^{-6}-3.0\times10^{-4}{\tan^2\beta}+{4.3\times10^{-8}}{\tan^4\beta},&\text{type-II},\\
\displaystyle 1+5.5\times10^{-5}-\frac{2.5\times10^{-4}}{\tan^2\beta}+\frac{3.0\times10^{-6}}{\tan^4\beta},&\text{type-X},\\
\displaystyle 1+3.1\times10^{-4},&\text{type-Y}.
\end{cases}
\end{align*}
Here, both the charged and the neutral Higgs bosons are involved, which results in the following features:
\begin{itemize}
\item  In all the four types, the 2HDM effects are strongly suppressed by the large mass of CP-even Higgs $m_H$ and small leptonic Yukawa coupling, and could be enhanced by the small mass of CP-odd Higgs $m_A$.
\item  In the type-II (-I and -X) models, the suppressed 2HDM contributions can be compensated by large $\tan\beta$ ($\cot\beta$).
\item  In the type-Y model, the 2HDM effect is $\beta$-independent. However, due to the large suppression, it can not provide strong bound on the masses of the Higgs bosons.
\end{itemize}

Under the constraints from $\overline{\mathcal B} (B_s\to \mu^+\mu^-)$ and $\overline{\mathcal B} (B_d \to \mu^+ \mu^-)$, the allowed parameter space $(m_H,m_A,m_{H^\pm},\tan\beta)$ of the four types of 2HDM are obtained, which are plotted in the $\tan\beta-m_{H^\pm}$ plane in figure~\ref{resultofbstomumu}. Due to the large error bars, the current experimental data put almost no constraint on the model parameters, except for the small excluded regions in the type-I and -II 2HDMs.

%
\subsection{Combined analysis and discrimination between the four 2HDMs}
\label{sec:combined constraints}

\begin{figure}
\centering
\includegraphics[width=0.43\textwidth]{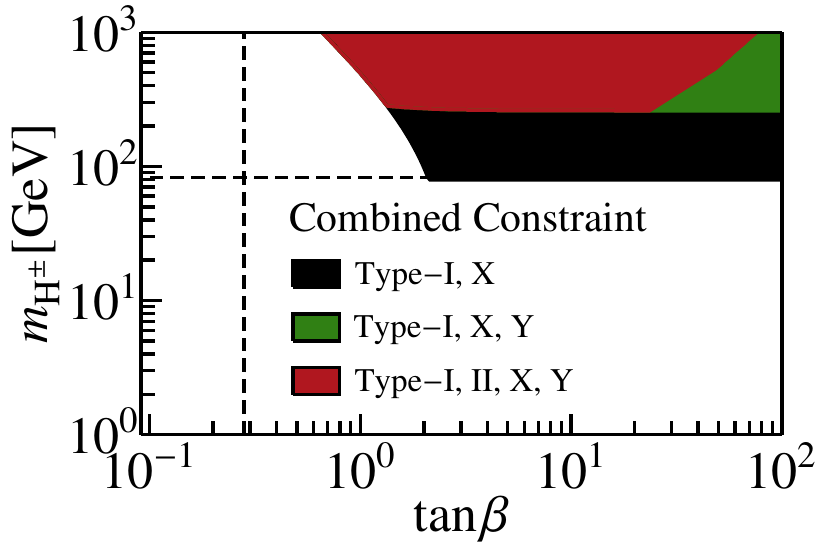}
\caption{\small Combined constraints on the parameter space of the four types of 2HDM, plotted in the $\tan\beta-m_{H^\pm}$ plane. The \textit{horizontal dashed line} denotes the direct bound on $m_{H^\pm}$ from the LEP experiment. The \textit{vertical dashed line} denotes the bound on $\tan\beta$ from perturbative unitarity and vacuum stability.}
\label{allchannelalltype}
\end{figure}

Combining all the constraints mentioned in the previous sections, we obtain the surviving parameter space as shown in figure~\ref{allchannelalltype}. From this plot, the following observations are made:
\begin{itemize}
\item For small $\tan\beta$, the most stringent constraints come from $\Delta m_{B_s}$ and $\mathcal B (\bar B \to X_s \gamma)$ in all the four types of 2HDM.
\item For large $\tan\beta$, the flavour observables put almost no constraints in the type-I and -X models. The LEP bound on $m_{H^\pm}$ is still the most strongest. For type-II and -Y models, the constraints mainly come from $\mathcal B (\bar B \to X_s \gamma)$. The $B_{s,d} \to \mu^+ \mu^-$ decays exclude one additional parameter space of the type-II 2HDM.
\item When $m_{H^\pm}$ become large, the combined constraints from flavour observables are almost the same for all the four 2HDMs.
\item The allowed region of the type-II model is contained in the one of the type-Y model, which stay in the survived parameter space of the type-I and -X 2HDMs. Therefore, the type-II model can be distinguished from the other 2HDMs in the green region in figure~\ref{allchannelalltype}, and the type-II and -Y models from the other 2HDMs in the black region.
\end{itemize}

\subsection{Other observables in $B_{s,d} \to \mu^+\mu^-$ within 2HDM}
\begin{figure}
\centering
    \subfigure[]{
    \label{resdeltagamma}
    \includegraphics[width=0.43\textwidth]{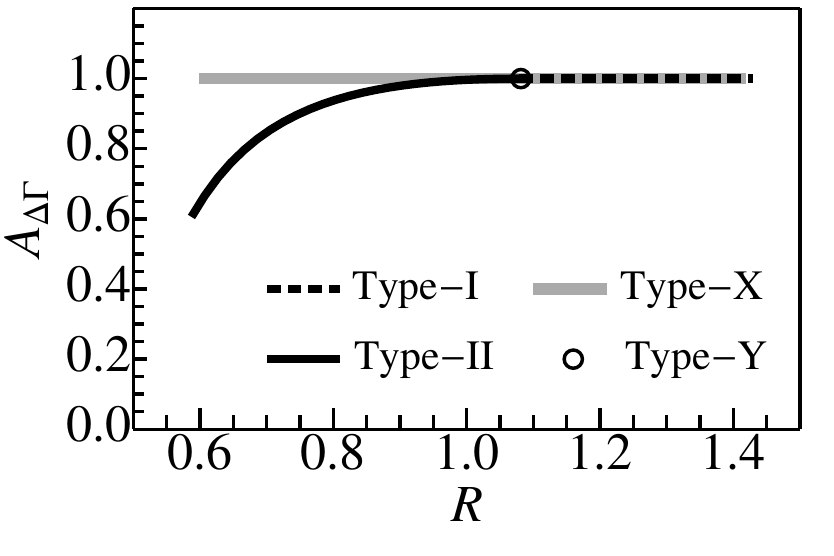}}
    \qquad
    \subfigure[]{
    \label{ratioofbstomumuandbdtomumu}
    \includegraphics[width=0.43\textwidth]{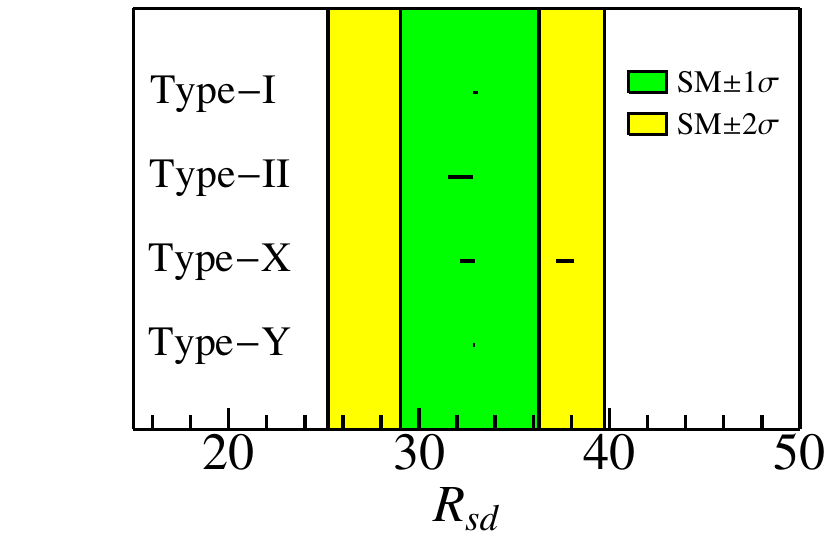}}
  \caption{\small (a) Correlations in the $R-A_{\Delta\Gamma}$ plane and (b) predicted ranges of $R_{sd}$ in the four types of 2HDM.}
\end{figure}

At present, only the branching ratio of $B_{s,d}\to \mu^+\mu^-$ have been measured. For the observables $A_{\Delta\Gamma}$ and $R$ defined in eq.~\eqref{eq:adeltgamma} and \eqref{eq:ratioofexpandsm}, we show in figure~\ref{resdeltagamma} the correlations between them within the four types of 2HDM, which are obtained in the parameter space given in section~\ref{sec:combined constraints}. The 2HDM predictions on $R_{sd}$ defined in eq.~\eqref{Eq:ratiodefine} are also shown in figure~\ref{ratioofbstomumuandbdtomumu}.  From these plots, we make the following observations:
\begin{itemize}
\item For the type-II 2HDM, large derivations from the SM predictions for both $A_{\Delta\Gamma}$ and $R$ are allowed, since both the Wilson coefficients $C_S$ and $C_P$ can be significantly enhanced by large $\tan\beta$. It is also noted that the observable $R$ always decreases.
\item For the type-I and -X 2HDMs, only large regions for $R$ are allowed. The reason is that the coefficient $C_S$ can not be enhanced by small $\tan\beta$, which has been excluded by the combined constraints discussed in section~\ref{sec:combined constraints}.
\item For the type-Y 2HDM, as expected, its effects on both $A_{\Delta\Gamma}$ and $R$ are small.
\item The observables $A_{\Delta\Gamma}$ and $R$ show the potential to
  discriminate  the four types of 2HDM. For the type-I, -II and -X
  models, there always exists an allowed region for only one of them in the $R-A_{\Delta\Gamma}$ plane. Interestingly, the allowed region of the type-Y is located in the intersection of  the regions of the other 2HDMs. With refined measurement of $A_{\Delta\Gamma}$ and $R$, one could distinguish one type 2HDM from others, or exclude all the four types.
\item At present, due to large uncertainties, the observable $R_{sd}$ can not provide further information to distinguish between the four 2HDMs. It is also noted that $R_{sd}$ always decreases in the type-II 2HDM.
\end{itemize}

It is concluded that the observables $A_{\Delta\Gamma}$, $R$ and $R_{sd}$ in $B_{s,d}\to \mu^+ \mu^-$ decays show high sensitivity to the Yukawa structure of the 2HDMs. Improved experimental measurements and theoretical predictions  will make these observables more powerful to distinguish between the four types of 2HDM.
\section{Conclusions}
\label{sec:conclusion}
In this paper, we have studied the possibility to discriminate the four types of 2HDM in the light of recent flavour physics data, including the $B_s-\bar B_s$ mixing, the leptonic B-meson decays $B_{s,d}\to \mu^+\mu^-$ and $B\to \tau \nu$, and the inclusive radiative decay $\bar B \to X_s \gamma$, together with the experimental data from the direct search for Higgs bosons at LEP, Tevatron and LHC~\cite{Heister:2002ev,ATLAS:1,ATLAS:2,CMS:1,CMS:2,Aaltonen:2012qt} and the constraints from perturbative unitarity~\cite{Ref:LQT} and vacuum stability~\cite{Sher:1988mj}. The outcomes of this combined analysis are summarized as follows:
\begin{itemize}
\item The flavour observables exhibit different dependence on the Yukawa couplings in the four types of 2HDMs. With the current experimental data, the allowed region of the type-II model is contained in the one of the type-Y model, which stay in the survived parameter space of the type-I and -X 2HDMs.
\item The observables $A_{\Delta\Gamma}$ and $R$ in the $B_s \to \mu^+ \mu^-$ decay, which arise from  the sizable $B_s$ width difference, are investigated. The correlation between these two observables is found to be sensitive probe to the Yukawa structure of 2HDM.
\end{itemize}

With the experimental progress expected from the LHC and the future
SuperKEKB, as well as the theoretical improvements, the constraints
shown here are expected to be refined,  which are helpful to
discriminate  the Yukawa structure if an extended Higgs sector is discovered in the future.

\section*{Acknoledgements}
The work was supported by the National Natural Science Foundation of China (NSFC) under contract Nos. 11225523 and 11221504. X. D. Cheng was also supported by the CCNU-QLPL Innovation Fund (QLPL201308). X. B. Yuan was also supported by the CCNU-QLPL Innovation Fund (QLPL2011P01) and the Excellent Doctoral Dissertation Cultivation Grant from Central China Normal University.

\begin{appendix}
\section{The Inami-Lim function $S(x_t,x_{H^\pm})$}
\label{appendix:1}
Within the four types of 2HDM discussed in section~\ref{sec:formalism}, the Inami-Lim function appearing in $B_s-\bar B_s$ mixing is given as~\cite{Urban:1997gw}
\begin{align}
S(x_t, x_{H^\pm})=S_{WW}(x_t)+2S_{WH}(x_t, x_{H^\pm})+S_{HH}(x_t, x_{H^\pm}),\label{Eq:s2HDM}
\end{align}
where the basic functions $S_{WW}$ and $S_{WH,HH}$ correspond to the $W$ box and the Higgs box diagrams shown in figure~\ref{LOboxen}, respectively. For convenience, their explicit expressions are given here:
\begin{flalign}
    &S_{WW}(x_t)&\nonumber\\
=&+\left[ 1-\frac{11x_t}{4}+\frac{x_t^2}{4}-\frac{3x_t^2\ln x_t}{2 ( 1-x_t )} \right] \frac{x_t}{( 1-x_t )^2}, &
\end{flalign}
\begin{flalign}
    &2S_{WH}(x_t,x_{H^\pm})& \nonumber\\
=&+\bigg[ \frac{x_t^2 x_{H^\pm} ( x_{H^\pm}-4 )}{2( 1-x_{H^\pm} )( x_t-x_{H^\pm} )^2}\ln \frac{x_t}{x_{H^\pm}} +\frac{3x_t^2\ln x_t}{2{{( 1-x_t )}^2}( 1-x_{H^\pm} )} -\frac{x_t^2( 4-x_t )}{2( x_{H^\pm}-x_t )( 1-x_t )} \bigg]( \xi_A^u )^2&\nonumber\\
& +\left[ \frac{x_t^2}{( x_{H^\pm}-x_t )^2} \right.\ln\frac{x_{H^\pm}}{x_t} +\left.\frac{x_t}{( x_t-x_{H^\pm} )} \right]  \times \left( \frac{1}{12}-\frac{m_{B_s}^2}{2 \tilde m_{B_s} ^2} \right) 3\sqrt{x_b x_s} (\xi _A^d )^2, &
\end{flalign}
\begin{flalign}
    &4S_{HH}(x_t,x_{H^\pm})& \nonumber\\
=& +\left[\frac{x_{H^\pm}+x_t}{(x_t-x_{H^\pm})^2} +\frac{2x_t x_{H^\pm} }{(x_{H^\pm}-x_t)^3} \ln\frac{x_t}{x_{H^\pm}} \right]x_t^2{(\xi _A^u)^4}& \nonumber\\
&  +\left[\frac{x_t^2+x_t x_{H^\pm}}{( x_{H^\pm}-x_t)^2x_{H^\pm}}+\frac{2x_t^2} {(x_{H^\pm}-x_t)^3} \ln \frac{x_t}{x_{H^\pm}}\right]x_b x_s (\xi _A^d)^4& \nonumber\\
&  +\left[ \frac{2}{(x_{H^\pm}-x_t)^2}+\frac{x_t+ x_{H^\pm} }{(x_{H^\pm}-x_t)^3}  \ln\frac{x_t}{x_{H^\pm}} \right]\left( -\frac{3}{2}+\frac{m_{B_s}^2}{\tilde m_{B_s}^2} \right) x_t^2\sqrt{x_b x_s} ( \xi_A^u\xi_A^d)^2& \nonumber\\
&   +\left[\frac{2}{(x_{H^\pm}-x_t)^2}+\frac{x_t+x_{H^\pm}}{( x_{H^\pm}-x_t )^3}\right] \left[(x_b+x_s)\frac{5m_{B_s}^2}{2 \tilde m_{B_s}^2}+\sqrt{x_b x_s} \left(1 - \frac{6m_{B_s}^2}{ \tilde m_{B_s}^2} \right) \right]x_t^2 ( \xi_A^u\xi_A^d)^2,&
\end{flalign}
with $\tilde m_{B_s}^2\equiv (\overline m_b(\overline m_b)+\overline m_s(\overline m_b))^2$ and $x_q \equiv (\overline m_q (\overline m_b))^2/m_W^2$ for $q=s,b$.
\section{The Wilson coefficients $C_S$ and $C_P$}
\label{appendix:2}
Within the four types of 2HDM discussed in section~\ref{sec:formalism}, the Wilson coefficients $C_S$ and $C_P$ appearing in the effective Hamiltonian of $B_{s,d} \to \mu^+ \mu^-$ are given as~\cite{Logan:2000iv}
\begin{align}
C_S&=C_S^{\rm box}+C_S^{\rm peng}+C_S^{\rm self}, \nonumber\\
C_P&=C_P^{\rm box}+C_P^{\rm peng}+C_P^{\rm self},
\end{align}
where the functions $C_{S,P}^{\rm box}$, $C_{S,P}^{\rm peng}$ and $C_{S,P}^{\rm self}$ correspond to the box, penguin and self-energy
diagrams associated with Higgs bosons in figure~\ref{btomumusm}. Based on the results in ref.~\cite{Logan:2000iv}, their explicit expressions are given here:
\begin{align}
C_{S,P}^{\rm box}&=\frac{m_\mu}{2} \frac{\xi _A^\ell \xi_A^d} {m_W^2}B_+(x_{H^\pm},x_t), \nonumber\\
C_S^{\rm peng}&=\frac{m_\mu}{2} \left[ \frac{\cos(\alpha-\beta)\xi _h^\ell \xi_A^{d}}{m_h^2} (1 - x_{H^\pm} + x_h) - \frac{\sin (\beta -\alpha)\xi_H^\ell\xi_A^d}{m_H^2} \left (1 - x_{H^\pm} + x_H \right) \right] P_+(x_{H^\pm} , x_t) \nonumber \\
C_P^{\rm peng} &=\frac{m_\mu}{2}\frac{\xi_A^\ell \xi_A^{d}}{m_{A}^2} (1-x_{H^\pm}+x_A) P_+(x_{H^\pm}, x_t), \nonumber\\
C_S^{\rm self} &=\frac{m_\mu}{2} \xi_A^u \xi_A^d \left( \frac{\xi_H^\ell \xi_H^d} {m_H^2}+\frac{\xi_h^\ell \xi _h^d}{m_h^2} \right)  (x_{H^\pm}-1) P_+(x_{H^\pm} , x_t) , \nonumber\\
C_P^{\rm self} &=\frac{m_\mu}{2} \xi _A^u \xi _A^{d} \left( \frac{\xi _A^\ell \xi_A^d}{m_A^2} \right)  (x_{H^\pm }-1) P_+(x_{H^\pm} , x_t),
\end{align}
with the functions
\begin{align}
B_+(x, y)&=\frac{y}{x-y}\left(\frac{\ln y}{y-1} -\frac{\ln x}{x-1}\right), \nonumber\\
P_+(x, y)&=\frac{y}{x-y}\left( \frac{x\ln x}{x-1}-\frac{y\ln y}{y-1} \right).
\end{align}
\end{appendix}

\end{document}